\documentclass[letter]{aa}  
\nolinenumbers
\let\linenumbers\relax
\let\nolinenumbers\relax
\usepackage{graphicx}
\usepackage{txfonts}
\usepackage{lscape}             
\usepackage{placeins}           

\usepackage{amsmath}
\usepackage{xcolor}
\usepackage{comment}
\usepackage[colorlinks=true,
  linkcolor=magenta,   
  citecolor=blue,  
  filecolor=magenta,
  urlcolor=magenta,    
  ]{hyperref}
\usepackage{amsfonts}

\usepackage{supertabular}
\usepackage{tabularx}
\usepackage{booktabs}

\begin{document}

   \title{An upper limit on Galactic OB+black hole binaries from SDSS-V}

%
%
%

    \author{J. M\"uller-Horn\inst{\ref{inst:mpia}, \ref{inst:uni-hd}}\corrauth{mueller-horn@mpia.de}        
        \and H.-W. Rix \inst{\ref{inst:mpia}} 
        \and S. Shahaf \inst{\ref{inst:mpia}}
        \and B. Pennell \inst{\ref{inst:mpia}, \ref{inst:uni-hd}}
        \and E. Zari \inst{\ref{inst:firenze}}
        \and M.C. Ram\'irez-Tannus \inst{\ref{inst:mpia}} 
        \and A. Farah \inst{\ref{inst:cita}}
        \and R. Essick \inst{\ref{inst:cita}}
        \and J.I. Villase\~nor \inst{\ref{inst:mpia}}
        \and M.R. Blanton \inst{\ref{inst:carnegie}}
        \and J.G. Fern\'andez-Trincado \inst{\ref{inst:santiago}}
        \and J.A. Kollmeier \inst{\ref{inst:carnegie}}
        \and C. Lam \inst{\ref{inst:carnegie}}
        \and K. Maltsev \inst{\ref{inst:mpia}}
        \and A. Roman-Lopes \inst{\ref{inst:laserena}}
        \and D.P. Schneider \inst{\ref{inst:penn}}
        \and A. Tkachenko \inst{\ref{inst:KUL}}
        }

    \institute{Max-Planck-Institut für Astronomie, Königstuhl 17, 69117 Heidelberg, Germany \label{inst:mpia}
    \and {Fakultät für Physik und Astronomie, Universität Heidelberg, Im Neuenheimer Feld 226, 69120 Heidelberg, Germany \label{inst:uni-hd}}
    \and {INAF–OAA, Osservatorio Astrofisico di Arcetri, largo E. Fermi 5, 50127, Firenze, Italy \label{inst:firenze}}
    \and {Canadian Institute for Theoretical Astrophysics, University of Toronto, 60 St. George Street, Toronto, ON M5S 3H8 \label{inst:cita}}
    \and {The Observatories of the Carnegie Institution for Science, 813 Santa Barbara Street, Pasadena, CA 91101, USA \label{inst:carnegie}}
    \and {Centro de investigaci\'on en Astronom\'ia, Facultad de Ingenier\'ia, Ciencia y Tecnolog\'ia, Universidad Bernardo O’Higgins, Av. Viel 1497, Santiago, 8370993, Chile \label{inst:santiago}}
    \and {Department of Astronomy, Universidad de La Serena, Av. Raul Bitran 1302, La Serena, Chile \label{inst:laserena}}
    \and {Department of Astronomy and Astrophysics, The Pennsylvania State University, University Park, State College PA16802, USA \label{inst:penn}}
    \and {Institute of Astronomy, KU Leuven, Celestijnenlaan 200D, 3001 Leuven, Belgium \label{inst:KUL}}
    }
   \date{Received October 12, 2026}

 \abstract
{Binaries composed of a luminous OB star and a black hole (BH) are a key evolutionary link between massive stellar binaries and merging compact-object binaries, but only a handful are known and their occurrence fraction remains poorly constrained.
We use multi-epoch spectroscopy from the SDSS-V Milky Way Mapper to place a first rigorous empirical upper limit on the fraction of Galactic massive (${\geq} 10 \, \mathrm{M_\odot}$) OB stars with a close BH companion, $f_{\rm BH}$.
Radial-velocity (RV) variability alone cannot distinguish BH companions from the far more numerous luminous binaries. We therefore combine the SDSS-V RV variability measurements of 3130 massive OB stars with high-resolution follow-up and literature information to identify luminous companions among the RV-variable subset. 
Forward-modelling the detectability of OB+BH systems using BH masses and orbital parameters from binary-evolution models, we find  $f_{\rm BH}\lesssim 3\%$ at 95\% credibility. 
This empirical upper limit approaches, but does not exclude, theoretical predictions for the OB+BH fraction.
The constraint is primarily limited by the number of RV-variable systems that can be decisively classified, highlighting the importance of complete, high-quality follow-up for future searches.
}


   \maketitle

\section{Introduction}

The mergers of double compact-object (CO) binaries observed through gravitational waves provide a unique probe of massive binary evolution, but the evolutionary pathways leading to these systems remain uncertain. A substantial fraction of merging neutron-star and black-hole (BH) binaries may form through the isolated evolution of massive stellar binaries, making the intermediate stages between massive stellar binaries and double-CO systems important observational benchmarks \citep[e.g.][]{Mapelli2021, Mandel_Broekgaarden2022}.
Binaries containing a luminous OB star and a BH companion represent one such evolutionary stage. These systems have survived the formation of the first CO while remaining bound, and their occurrence fraction is sensitive to uncertain aspects of binary evolution, including mass transfer, stellar winds, and supernova physics. Binary-evolution calculations predict that about $1$--$4\%$ of present-day OB stars may host a BH companion, with some dependence on stellar mass \citep{Langer+2020}. However, only a handful of OB+BH systems are currently known, and their heterogeneous discovery channels hinder a robust empirical determination of their population frequency \citep{Shenar+2022, Mahy+2022}.
A statistical constraint requires a large, well-characterised sample of massive stars and a model for the detectability of BH companions. Radial-velocity (RV) variability provides a sensitive probe of an unseen companion, but sparse temporal sampling makes its detectability dependent on both the binary orbit and observing cadence. The analysis must therefore account for this selection function while distinguishing BH companions from the much more numerous population of ordinary stellar binaries.
The fifth generation of the Sloan Digital Sky Survey \citep[SDSS-V;][]{Kollmeier+2026,SDSS_DR20} provides multi-epoch spectroscopy for several thousand massive stars in the Milky Way, enabling such an analysis. In this Letter, we combine the observed RV variability of the SDSS-V OB-star sample with high-resolution spectroscopic follow-up and literature data to constrain the occurrence fraction $f_{\rm BH}$ of massive OB stars with BH companions.


\section{Data}
\label{sec:data}
\subsection{The SDSS-V census of Galactic OB stars}
\label{sec:sdss}

SDSS-V is obtaining spectra for several hundred thousand Galactic O-, B-, and A-type stars, substantially expanding the spectroscopic census of young and luminous stars in the Milky Way. The sample was selected according to a well-defined selection function \citep{Zari+2021} based primarily on photometric criteria in \textit{Gaia} and 2MASS.
Most targets were observed with the low-resolution BOSS spectrographs \citep{Smee+2013, Gunn+2006, Bowen_Vaughan1973}, covering 362--1035\,nm\ at a resolving power of $R {\simeq} 1800$. The brightest stars exceed the BOSS saturation limit and were instead observed with the high-resolution FEROS spectrograph, which covers 350--920\,nm at $R{\simeq}\,5 {\times} 10^4$ \citep[and a small subset with the FIES and HERMES instruments;][]{Kaufer+1999, Telting+2014, Raskin+2011}. We incorporated these observations into the same SDSS-V hot-star catalogue. Stellar parameters for the combined BOSS and high-resolution sample were derived homogeneously from a joint analysis of the optical spectra and flux-calibrated \textit{Gaia} spectrophotometry (Rix et al. 2026, in prep.), including stellar masses from isochrone fitting.
We selected a parent sample of massive stars ($M_\mathrm{OB} {\geq} 10\,\mathrm{M_\odot}$) with at least two epochs 
to enable a measurement of RV variability. The resulting sample contains $N {=} 3130$ stars, including 2653 observed with BOSS and 477 with FEROS, FIES and HERMES. 

\subsection{RV variability}
\label{sec:rv_variability}
\begin{figure}[t]
\centering
\includegraphics[width=\columnwidth]{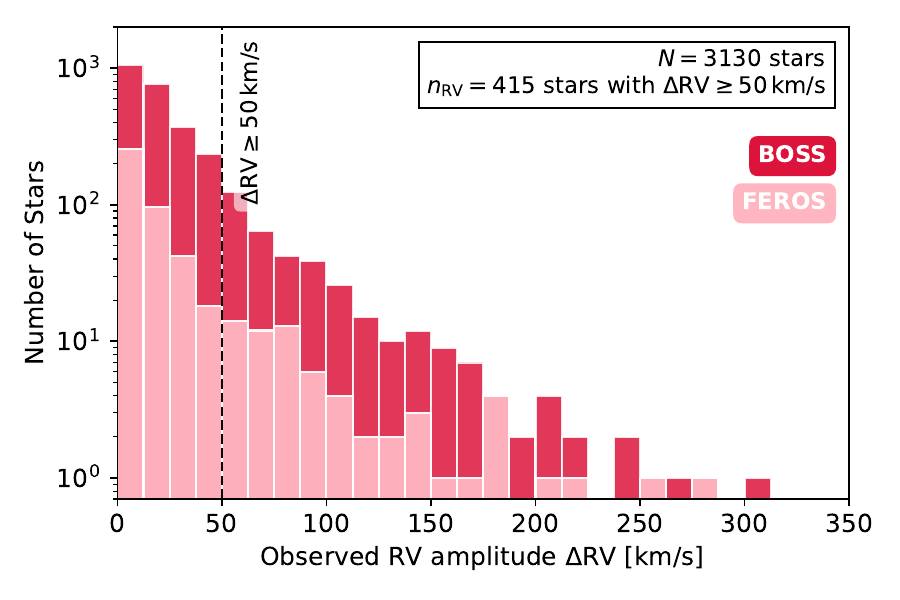}
\caption{Observed maximum RV differences in the OB parent sample. 
}
\label{fig:drv_histogram}
\end{figure}

For each star, we retrieved the observing epochs and corresponding RVs from the SDSS-V database. The epoch RVs were derived from spectral-line fitting of the hydrogen Balmer lines, with median uncertainties of 7 and 4$\,\mathrm{km\,s^{-1}}$ for BOSS and FEROS spectra, respectively. The stars have a median of three epochs and a median temporal baseline of about two months.
The sparse sampling generally prevents fitting full orbital solutions. We therefore characterised RV variability by the maximum observed difference
\begin{equation}
\Delta\mathrm{RV} = \mathrm{RV}_{\rm max}-\mathrm{RV}_{\rm min}. \label{eq:Delta_RV}
\end{equation}
Figure~\ref{fig:drv_histogram} shows the distribution of $\Delta\mathrm{RV}$ in the parent sample. 

Rather than modelling the full $\Delta\mathrm{RV}$ distribution, we focussed on the high-amplitude RV tail using the variability indicator,
\begin{equation}
Y_i =
\begin{cases}
1, & \mathrm{if}\ \Delta\mathrm{RV}_i > 50\,\mathrm{km\,s^{-1}}; \quad \\
0, & \text{otherwise}.
\end{cases} \label{eq:rv_criterion}
\end{equation}
The threshold focuses the analysis on systems with substantial observed RV variations, for which the relative contribution of OB+BH systems is expected to be higher and spectroscopic classification is more effective. The low-$\Delta\mathrm{RV}$ regime is more difficult to model reliably because of measurement uncertainties and other sources of low-amplitude variability. Applying this criterion identifies $n_\mathrm{RV}{=}415$ RV-variable stars, which may have luminous or compact binary companions. 

\subsection{Follow-up data from FEROS spectra and the literature}
\label{sec:follow_up}
Of the $n_{\rm RV}$ RV-variable stars, $m_\mathrm{fu} {=} 66$ were observed with FEROS; the remaining 349 have BOSS observations only. The high-resolution FEROS spectra can reveal luminous companions through, for example, a second set of spectral lines identifying a double-lined spectroscopic binary (SB2). For some stars, independent evidence for a luminous companion is also available in the literature, such as eclipses or dynamical companion-mass constraints. 
We combined the FEROS spectroscopy and literature information to classify the companions of these $m_\mathrm{fu}$ RV-variable stars. We assigned each star a companion-classification label, which we collectively refer to as the \emph{follow-up data} $\{D_j\}$, with $j{=}1,\ldots,m_{\rm fu}$. In total, 56 systems were classified as SB2s, eclipsing, or other contaminants that exclude an OB+BH interpretation, while no luminous companion was detected for the remaining ten systems. Details of the classification procedure and individual classifications are given in Appendix~\ref{app:follow_up}.

\section{Hierarchical inference of the OB+BH fraction}
\label{sec:model}

We combined the RV-variability information for the full parent sample with the companion constraints from the high-resolution follow-up in a hierarchical model to infer $f_{\rm BH}$, following the general framework for combining population-level data with follow-up observations described by \citet{Essick_Farah2026}. 
We define $f_{\rm BH}$ as the fraction of OB stars in the parent sample whose close companion ($P_\mathrm{orb} {\lesssim} 3\,$yr) is a BH.
The large parent sample constrains the overall fraction of RV-variable stars, while the follow-up data determine how much of this population can be attributed to luminous binaries and other contaminants rather than OB+BH systems. Because BH companions contribute essentially no optical light, the SDSS-V photometric selection is effectively independent of the presence of a BH companion; the parent-sample selection function therefore does not enter the inference of $f_{\rm BH}$.


\subsection{RV variability model}
\label{sec:likelihood_rv}

An RV-variable star ($Y_i {=} 1$) can be an OB+BH binary, a stellar binary, or an effectively single star whose measured RV amplitude is spuriously large, for example because of pulsations. We write the probability that star $i$ is RV-variable as
\begin{align}
P(Y_i{=}1|f_\mathrm{BH},\kappa_s,M_{\mathrm{OB},i})
&=
f_{\rm BH}\,P_{{\rm det},i}^{\rm BH}
+
f_{\rm bin}\,P_{{\rm det},i}^{\rm bin}
+
f_{\rm single}\,P_{{\rm det},i}^{\rm single}
\label{eq:theta_full} \\  
&\approx f_{\rm BH}\,P_{{\rm det},i}^{\rm BH}
+
\kappa_{s(i)}, 
\label{eq:theta}
\end{align}

\noindent where $f_{\rm BH/bin/single}$ are the corresponding occurrence fractions in the population and $f_{\rm BH} {+} f_{\rm bin} {+} f_{\rm single} {=} 1$. The detection probabilities $P_{{\rm det},i}^{\rm BH/bin/single}$ describe the probability that a system of the respective type satisfies the RV-variability criterion Eq.~\eqref{eq:rv_criterion}; see Appendix~\ref{app:bh_model}. Single stars are included for completeness but $P_{{\rm det},i}^{\rm single}$ and their contribution to the RV-variable sample are expected to be small.
We explicitly modelled $P_{{\rm det},i}^{\rm BH}$ on a star-by-star basis. For each star, we simulated OB+BH binaries using its inferred primary-star mass and actual spectroscopic observing epochs. We drew BH masses and orbital parameters from the distributions predicted by \citet{Langer+2020}, calculated the RVs at the observed epochs, and defined $P_{{\rm det},i}^{\rm BH}$ as the fraction of simulated systems satisfying Eq.~\eqref{eq:rv_criterion}. Details of the adopted BH population model and the calculation of these detection probabilities are given in Appendix~\ref{app:bh_model}.

In contrast, we did not explicitly model the underlying populations of stellar binaries or single-star RV variables. Instead, their aggregate contribution was absorbed into $\kappa_{s(i)}$. Because the probabilities depend on the temporal sampling, we divided the stars into three bins according to their number of spectroscopic epochs: two, three, and four or more. We approximated the combined non-BH contribution within each bin by a common effective value $\kappa_s$, with $s(i)$ denoting the observing-epoch bin of star $i$. This stratification captures the dominant dependence on temporal sampling while avoiding the need to specify explicit models for the stellar-binary and single-star populations. (For a test of the sensitivity of our results to this assumption, see Appendix~\ref{app:robustness_fbin-model}.)
Assuming independent Bernoulli trials, the likelihood of the RV-variability indicators is
\begin{multline}
P(\{Y_i\}\mid f_{\rm BH},\kappa_s) =
\prod_{i=1}^{n_\mathrm{RV}}
P(Y_i{=}1|f_\mathrm{BH},\kappa_s,M_{\mathrm{OB},i})
\\ \times \prod_{j=1}^{N-n_\mathrm{RV}}
[1 - P(Y_j{=}1|f_\mathrm{BH},\kappa_s,M_{\mathrm{OB},j})]
\,.
\label{eq:likelihood_det}
\end{multline}

\subsection{Follow-up constraints}
\label{sec:likelihood_fu}
For the $m_\mathrm{fu}$ RV-variable stars with FEROS spectroscopy and additional literature information, the follow-up data $\{D_k\}$ provide independent constraints on the nature of the companion. We encode this information as
$
\phi_k {=} P({\rm BH}\mid D_k),
$
computed assuming equal prior probabilities for the BH and non-BH hypotheses. Thus, $\phi_k {=} 0$ corresponds to a companion classification that excludes an OB+BH interpretation, whereas $\phi_k {=} 0.5$ indicates that the available data provide no constraint, and $\phi_k {=} 1$ a secure OB+BH detection.

For any star $k$ that is RV-variable ($Y_k = 1$), the probability that it belongs to the OB+BH population can be written as 
\begin{align}
\pi_{{\rm BH},k}
=
P({\rm BH}\mid Y_k{=}1,f_{\rm BH},\kappa_s)
& =
\frac{f_{\rm BH}P_{{\rm det},k}^{\rm BH}} 
{P(Y_k{=}1|f_\mathrm{BH},\kappa_s,M_{\mathrm{OB},k})}.
\label{eq:qbh}
\end{align}
\noindent Marginalising over the unknown companion class, the likelihood of the follow-up data becomes
\begin{multline}
P(\{D_k\} \,| \, \{Y_k{=}1\}, f_\mathrm{BH}, \kappa_{s}) 
\propto
\prod_{k=1}^{m_{\rm fu}}
\left[
\pi_{{\rm BH},k}\,\phi_k+
(1{-}\pi_{{\rm BH},k})\,(1{-}\phi_k)
\right],
\label{eq:likelihood_fu}
\end{multline}

\noindent where the proportionality constant does not depend on any parameters of interest.
A derivation of this expression is given in Appendix~\ref{app:follow_up-likelihood}.

\subsection{Posterior inference}
Our model is parametrised by the OB+BH occurrence fraction $f_{\rm BH}$ and the three effective contributions $\kappa_s$ of RV-variable stars in the different epoch-number bins. Combining the RV and follow-up likelihoods, the joint posterior is
\begin{multline}
p(f_{\rm BH},\kappa_{s}\mid\{Y_i\}, \{D_j\}) \\
\propto
p(f_{\rm BH},\kappa_{s})\,
p(\{Y_i\}|f_\mathrm{BH}, \kappa_s)\,
p(\{D_j\}|\{Y_j=1\}, f_\mathrm{BH}, \kappa_s)\,.
\end{multline}
\noindent We adopted a uniform prior $p(f_{\rm BH},\kappa)$ over $0 {\leq} f_{\rm BH} {\leq} 1$ and $0 {\leq} \kappa_{s} {\leq} (1 {-} f_{\rm BH})$ for each bin $s$. We sampled the posterior using the \texttt{numpyro} Python package's \texttt{NUTS} sampler. 

\section{Results}
\label{sec:results}
\begin{figure}[t]
   \centering
   \includegraphics[width=\columnwidth]{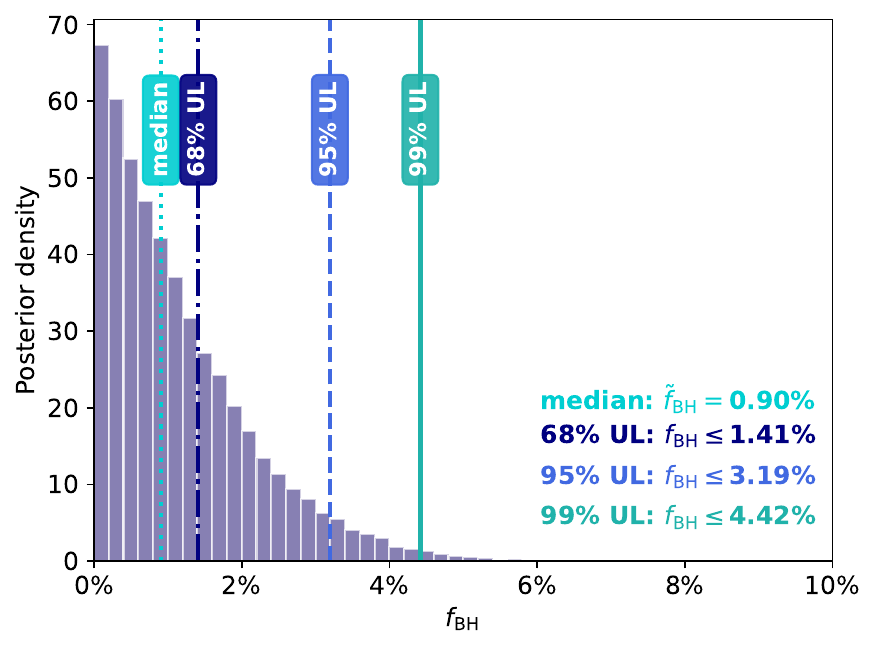}
   \caption{Normalised posterior probability density distribution of the OB+BH fraction $f_{\rm BH}$. Vertical lines indicate the median value and the $68\%$, $95\%$, and $99\%$ upper limits of $f_{\rm BH} \leq 1.41\%,$ $3.19\%,$ $\mathrm{and}\,4.42\%.$}
   \label{fig:fbh_posterior}
\end{figure}

Figure~\ref{fig:fbh_posterior} shows the posterior distribution of the OB+BH occurrence fraction $f_{\rm BH}$, marginalised over the nuisance parameters $\kappa_s$. The posterior peaks at $f_{\rm BH} {=} 0$, indicating that the data do not require a positive contribution from OB+BH binaries. We therefore interpret the upper quantiles of the posterior as upper limits on their occurrence fraction.
The $68\%$, $95\%$, and $99\%$ upper limits are
$
f_{\rm BH} \leq 1.41\%,$ $3.19\%,$ $\mathrm{and}\,4.42\%,$ respectively. Under the fiducial OB+BH population model, this corresponds to fewer than $\sim 3\%$ of the massive stars in our sample hosting a close BH companion at $95\%$ credibility.
Given the systematic uncertainties associated with the modelling assumptions explored in Appendix~\ref{app:robustness}, we report our main result as a $95\%$ upper limit of $f_{\rm BH}\lesssim 3\%$.
For reference, assuming a present-day Galactic population of $\sim1.2 \times10^5$ massive OB stars (estimated based on \citealt{Quintana+2026} and extrapolated following \citealt{Reed2005}), the $95\%$ upper limit corresponds to ${\lesssim} 4000$ OB+BH binaries currently present in the Milky Way, ${\lesssim} 100$ in our SDSS-V sample, and ${\lesssim} 60$ within 2\,kpc of the Sun.

\section{Discussion}
\label{sec:discussion}

\subsection{Model assumptions and comparison with predictions}
Our constraint on $f_{\rm BH}$ depends on the adopted OB+BH population model and on approximations in the hierarchical inference. We accounted for the detectability of OB+BH systems on a star-by-star basis using the inferred stellar masses and actual observing epochs, adopting the BH-mass and orbital-parameter distributions of \citet{Langer+2020}. The resulting constraint therefore applies to populations with broadly similar properties; systems concentrated at less readily detectable orbital periods or configurations would be less constrained. We tested the sensitivity to the assumed BH masses and orbital eccentricities in Appendix~\ref{app:robustness}, finding only modest changes to the inferred upper limit. 

We modelled the contaminating population of luminous stars and stellar binaries through the effective nuisance parameters $\kappa_s$, rather than explicitly modelling their heterogeneous orbital properties. This provides a flexible description of the RV-variable population without requiring an uncertain binary-population model. We tested alternative model set-ups in Appendix~\ref{app:robustness}. There we also tested the potential differences between the BOSS and FEROS samples by allowing independent $\kappa_s$ parameters for the two subsamples. The resulting constraints remain consistent with the fiducial model, indicating that these simplifications do not significantly affect our result.

The inferred occurrence fraction applies to the population represented by our SDSS-V sample and is therefore averaged over the sample's distribution of primary masses. \citet{Langer+2020} predict a mass-dependent OB+BH fraction of approximately $1$--$4\%$ over the relevant stellar-mass range. Applying their predicted mass-dependent fraction to our observed mass distribution gives an expected mean occurrence fraction of approximately $2.5\%$. Our 95\% upper limit of $3.19\%$ is comparable to this prediction but does not exclude it. 

\subsection{What determines the sensitivity of the experiment}
\label{sec:discussion_strategy}

The structure of our hierarchical inference provides guidance for future BH searches. The large SDSS-V sample constrains the overall fraction of RV-variable stars precisely, but this population contains both stellar binaries and potential OB+BH systems. The high-resolution follow-up is what breaks this degeneracy by identifying luminous companions and thereby constraining which systems can be attributed to unseen companions. The statistical power therefore comes from combining a precise measurement of the overall RV-variable fraction with informative follow-up classifications.

Our current constraint is primarily limited by the fraction of decisively classified RV-variable systems. In the absence of BH detections, the upper limit on $f_{\rm BH}$ scales approximately as $m_{\rm fu,eff}^{-1}$, the \emph{effective} number of follow-up targets, which can be less than the total number of classified systems. This is because inconclusive non-detections favour the BH hypothesis over the non-BH hypothesis, and therefore partially offset the constraining power of decisive non-BH classifications. The effective follow-up sample is consequently determined by both the number of targets and the completeness of their classification. Once BH companions are detected, the fractional precision on $f_\mathrm{BH}$ scales approximately as $k^{-1/2}$, where $k$ is the number of detections (Appendix~\ref{app:experiment_design}).

For future searches, longer temporal baselines and better orbital-phase coverage in the parent sample can improve the detectability of OB+BH systems, while additional high-resolution, multi-epoch follow-up observations help increase the number of RV-variable systems that can be decisively classified. A targeted follow-up sample with high classification completeness can provide substantially more statistical leverage than sparse classification of a larger sample. This provides a direct path towards tighter upper limits and, ultimately, a measurement of the OB+BH fraction.

\section{Conclusions}


We used multi-epoch spectroscopy from the SDSS-V hot-star census to constrain the occurrence fraction of Galactic OB stars with BH companions. Of $3130$ massive stars ($M_\mathrm{OB} {\geq} 10\,\mathrm{M}_\odot$), $415$ show RV amplitudes exceeding $50\,\mathrm{km\,s^{-1}}$. We modelled this RV-variable population as a mixture of OB+BH systems, luminous stars and stellar binaries, incorporating star-specific OB+BH detection probabilities and companion constraints from high-resolution spectroscopy and the literature.
Under the assumed OB+BH population model \citep{Langer+2020}, we found a $95\%$ upper limit of ${f_{\rm BH} {\lesssim} 3\%},$ corresponding to fewer than ${\lesssim} 4000$ OB+BH binaries in the Milky Way. This constraint approaches the ${\sim}2.5\%$ expected for our sample's mass distribution, without yet excluding this prediction. 
Our result provides an empirical benchmark for the number of binaries that have survived mass transfer and formation of the first BH.
Future multi-epoch spectroscopy, improved RV coverage, \textit{Gaia} DR4, and surveys of massive stars in the Magellanic Clouds may enable measurements of the OB+BH occurrence fraction across stellar mass, orbital period, and metallicity.

\begin{acknowledgements}
This research was supported in part by grant NSF PHY-2309135 to the Kavli Institute for Theoretical Physics (KITP). This research benefited from discussions at the 
``Stellar-Mass Black Holes at the Nexus of Optical, X-ray, and Gravitational Wave Surveys'' programme at the Kavli Institute of Theoretical Physics. J.M.-H. and H.-W.R. acknowledge support from the European Research Council for the ERC Advanced Grant [101054731]. 
Funding for the Sloan Digital Sky Survey V has been provided by the Alfred P. Sloan Foundation, the Heising-Simons Foundation, the National Science Foundation, and the Participating Institutions. SDSS acknowledges support and resources from the Center for High-Performance Computing at the University of Utah. SDSS telescopes are located at Apache Point Observatory, funded by the Astrophysical Research Consortium and operated by New Mexico State University, and at Las Campanas Observatory, operated by the Carnegie Institution for Science. The SDSS web site is \url{www.sdss.org}.
SDSS is managed by the Astrophysical Research Consortium for the Participating Institutions of the SDSS Collaboration, including the Carnegie Institution for Science, Chilean National Time Allocation Committee (CNTAC) ratified researchers, Caltech, the Gotham Participation Group, Harvard University, Heidelberg University, The Flatiron Institute, The Johns Hopkins University, L'Ecole polytechnique f\'{e}d\'{e}rale de Lausanne (EPFL), Leibniz-Institut f\"{u}r Astrophysik Potsdam (AIP), Max-Planck-Institut f\"{u}r Astronomie (MPIA Heidelberg), Max-Planck-Institut f\"{u}r Extraterrestrische Physik (MPE), Nanjing University, National Astronomical Observatories of China (NAOC), New Mexico State University, The Ohio State University, Pennsylvania State University, Smithsonian Astrophysical Observatory, Space Telescope Science Institute (STScI), the Stellar Astrophysics Participation Group, Universidad Nacional Aut\'{o}noma de M\'{e}xico, University of Arizona, University of Colorado Boulder, University of Illinois at Urbana-Champaign, University of Toronto, University of Utah, University of Virginia, Yale University, and Yunnan University.
\end{acknowledgements}

%
\bibliographystyle{aa} 
\bibliography{references} 

\begin{appendix}




\twocolumn
\section{Follow-up data and analysis}
\label{app:follow_up}

\setlength{\tabcolsep}{17pt}
\begin{table}[h]
\centering
\caption{Overview of the number of stars in the parent sample and the RV-variable subset observed with BOSS and FEROS.}
\label{tab:sample_overview}
\begin{tabular}{lrr}
\toprule
& Total & RV variable   \\
& & $\Delta \mathrm{RV}>50\,{\rm km/s}$\\
\midrule
All & $N = 3130$ & $n_\mathrm{RV} = 415$ \\
BOSS & $N_{\rm BOSS} = 2653$ &  $n_{\rm BOSS} = 349$ \\
FEROS & $N_{\rm FEROS} = 477$ & $n_{\rm FEROS}= 66$ \\
\bottomrule
\end{tabular}
\end{table}

\setlength{\tabcolsep}{6pt}

Table~\ref{tab:sample_overview} summarises the stellar sample used in this work and the numbers of RV-variable stars identified from the BOSS and FEROS observations. The follow-up sample comprises the $m_\mathrm{fu} {=} 66$ FEROS stars with $\Delta \mathrm{RV} {>} 50\,\mathrm{km\,s}^{-1}$. These stars are listed individually in Table~\ref{tab:follow_up}, together with their \textit{Gaia} \texttt{source\_id} and measured maximum RV difference. We investigated the nature of their companions using the available high-resolution spectra together with archival data and information from the literature. The resulting classifications provide the additional information needed to distinguish stellar binaries from potential OB+BH systems and thereby break the degeneracy between $f_{\rm BH}$ and $\kappa_s$ in the RV-variable sample.

We first searched the FEROS spectra for evidence of a luminous companion through the presence of a second set of spectral lines. Many systems showed clear double-lined features in a visual inspection. We complemented this initial assessment with the Fourier-domain method developed by Shahaf et al.~(submitted), which uses the line profile variability of isolated spectral lines to place calibrated constraints on the presence of a faint secondary spectrum. We applied this method to the \ion{He}{I} $\lambda\lambda4471$, 5876, and 6678 lines. For 26 of the 66 stars, the analysis provided clear evidence for a luminous secondary. These systems are classified as \texttt{SB2} in Table~\ref{tab:follow_up} and were excluded as OB+BH systems, corresponding to $\phi_j {=} 0$ in the hierarchical inference.

We then searched archival data and the literature for additional constraints on the nature of the companions. We identified 27 known eclipsing binaries (\texttt{EB}) among the follow-up sample. Eclipses demonstrate the presence of a non-compact companion and therefore exclude an OB+BH interpretation. These systems were identified through cross-matches with the \textit{Gaia} DR3 catalogue of eclipsing binaries \citep{Mowlavi+2023} and the TESS OBA-type Eclipsing Binaries Catalogue \citep{IJspeert+2021,IJspeert+2024}. We found nine systems, labelled \texttt{SB9} in Table~\ref{tab:follow_up}, listed in the 9th Catalogue of Spectroscopic Binary Orbits \citep{Pourbaix+2004} with published spectral classifications for their luminous companions.

For several further systems, individual literature references provided evidence for luminous companions. There is one X-ray binary with a neutron-star companion and one SB1 system for which the published orbital solution implies a low-mass companion \citep{Britavskiy+2023,Rosslowe_Crowther2015,Blex+2024,Mahy+2022,Barba+2026}.\footnote{\href{http://pacrowther.staff.shef.ac.uk/WRcat/}{http://pacrowther.staff.shef.ac.uk/WRcat/}} These systems were likewise excluded as OB+BH candidates.

In addition to these companion classifications, we identified five stars for which the large measured $\Delta \mathrm{RV}$ is likely spurious and does not reflect genuine orbital motion. Three are Be stars (\texttt{Be star}) with strong emission in the hydrogen Balmer and \ion{He}{I} lines, which biased the RVs measured from those lines and overestimated the amplitude. Two blue supergiants showed variable H$\alpha$ emission and line-profile variability (\texttt{LPV}) rather than centre-of-mass motion. In each case the intrinsic RV amplitude was below our threshold. 
We therefore excluded all five objects as OB+BH candidates and assigned $\phi_j {\simeq} 0$.

\begin{figure}[h]
\centering
\includegraphics[width=0.9\columnwidth]{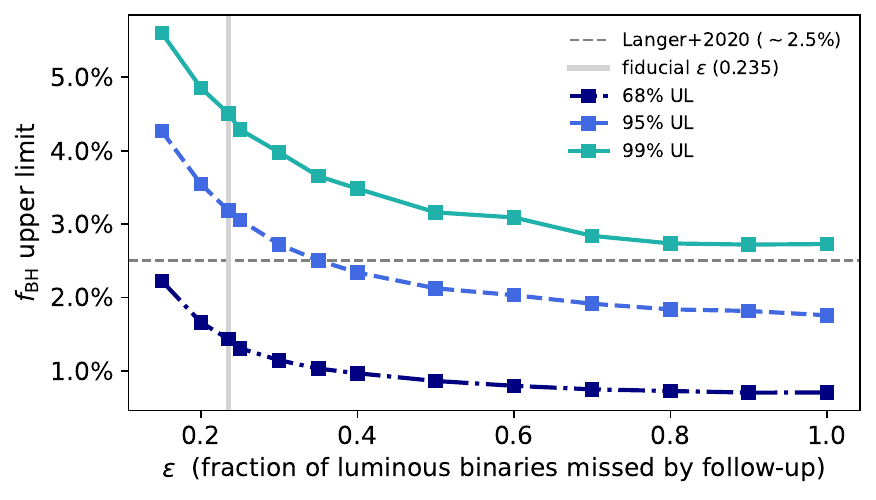}
\caption{Dependence of the inferred upper limit on $f_\mathrm{BH}$ on the assumed value of $\epsilon$; the fraction of RV-variable non-BH systems that evade our follow-up tests and classification. For the fiducial model, we adopted $\epsilon {\simeq} 0.235$ as a conservative lower estimate.}
\label{fig:eps_scan}
\end{figure}

Combining the spectroscopic analysis, archival information, and literature constraints, we excluded 56 of the 66 RV-variable FEROS stars as OB+BH candidates. 
For the remaining ten RV-variable stars, we did not identify a stellar companion through either a secondary spectrum or eclipses. Such a non-detection is consistent with a BH companion, but can also occur for non-BH companions that evaded our tests, for example because of a low flux ratio or unfavourable inclination. The non-detection is therefore informative but not decisive. We encoded this using the likelihood ratio in Eq.~\eqref{eq:likelihood_ratio}. Setting $P(D_j {\mid} {\rm BH}, Y_j {=} 1) {=} 1$ and $P(D_j {\mid} \neg{\rm BH}, Y_j {=} 1) {\equiv} \epsilon$, where $\epsilon$ is the fraction of RV-variable non-BH systems that escape our follow-up tests, gives
$
\phi_j {=} \frac{1}{1+\epsilon}.
$
The limit $\epsilon {=} 1$ corresponds to no classification sensitivity and recovers the uninformative value $\phi_j {=} 0.5$, while $\epsilon {\rightarrow} 0$ makes a non-detection increasingly indicative of a BH.
We estimated the follow-up incompleteness $\epsilon$ using a capture–recapture calculation, treating the spectroscopic and archival classifications as independent detection channels.
With $n_A$ systems identified spectroscopically, $n_B$ identified from archival sources, and $n_{AB}$ identified by both, the corresponding missed fractions are
$\epsilon_A {=} \frac{n_B-n_{AB}}{n_B},
\ 
\epsilon_B {=} \frac{n_A-n_{AB}}{n_A},$, and the joint missed fraction is $\epsilon {\simeq} \epsilon_A\epsilon_B$. For our sample, $n_A {=} 26$, $n_B {=} 41$, and $n_{AB} {=} 16$, yielding $\epsilon {\simeq} 0.235$ and hence $\phi_j {\simeq} 0.8$.
In reality, the two detection channels are not fully independent; both favour short-period, high-inclination, high-mass-ratio systems and are therefore positively correlated. This increases $n_{AB}$ relative to the independent case and causes the above estimate to underestimate $\epsilon$. We therefore adopt $\epsilon {\simeq} 0.235$ as a conservative lower estimate and show how our upper limits on $f_\mathrm{BH}$ change for higher $\epsilon$ as a robustness check in Fig.~\ref{fig:eps_scan}; for $\epsilon {\geq} 0.35$, the inferred upper limit drops below $f_\mathrm{BH}^{95} {\leq} 2.5\%$.
We note that, naively, we might have expected the constraints on $f_\mathrm{BH}$ to become weaker as $\epsilon$ increases because the follow-up observations would be less informative.
However, Fig.~\ref{fig:eps_scan} shows the opposite behavior.
We interpret this as due to the fact that we use the observations (Table~\ref{tab:follow_up}) for all tested values of $\epsilon$ in Fig.~\ref{fig:eps_scan}.
Because most systems with follow-up are confidently identified as not hosting BHs, high values of $\epsilon$ suggest that there must be a correspondingly large fraction of the systems that are not confidently identified as having a luminous companion but actually do have such a companion.
Because there are a fixed, finite number of such inconclusive stars in Table~\ref{tab:follow_up}, that in turn forces the inferred number of systems with BH to be lower and therefore $f_\mathrm{BH}$ decreases.




\vspace{0.4em}
\setlength{\tabcolsep}{4pt}
\begin{center}

\tablefirsthead{%
\toprule
Gaia DR3 &
$\Delta\mathrm{RV}$ &
$\phi$ &
Reference \\
\texttt{source\_id} &
[km\,s$^{-1}$] &
 &
 \\
\midrule
}

\tablehead{%
\toprule
Gaia DR3 &
$\Delta\mathrm{RV}$ &
$\phi$ &
Reference \\
\texttt{source\_id} &
[km\,s$^{-1}$] &
 &
 \\
\midrule
}

\tabletail{\midrule}

\tablelasttail{\bottomrule}

\topcaption{The follow-up sample of $m_\mathrm{fu} {=} 66$ stars with high-resolution spectra and high RV differences.}
\label{tab:follow_up}
\tiny
\begin{supertabular}{@{}lccr@{}}
5966509095207261056  & 277 & 0.0 &  \texttt{SB2},  \texttt{EB}   \\
4066278846098710016  & 254 & 0.0 &  \texttt{SB2},  \texttt{EB}   \\
5953634393855997824  & 225 & 0.0 &  \texttt{SB2},  \texttt{EB}   \\
5966550391303432576  & 210 & 0.0 &  \texttt{SB2},  \texttt{EB}   \\
5875491419647525248  & 180 & 0.0 &  \texttt{SB2},  \texttt{EB}, \texttt{SB9}   \\
4055916430035297792  & 178 & 0.0 &  \texttt{SB2}   \\
5350357519345171200  & 177 & 0.0 & literature (HD\,93162)  \\
5966667734115159040  & 175 & 0.0 &  \texttt{EB}   \\
4057914208312575360  & 172 & 0.0 &  \texttt{SB2},  \texttt{EB}   \\
5350370026290390912  & 150 & 0.0 & \texttt{SB9}   \\
5953632886347154304  & 147 & 0.0 &  \texttt{EB}   \\
5350357313186767104  & 142 & 0.0 &  \texttt{SB2}, \texttt{SB9}   \\
476270508301993344  & 139 & 0.0 &  \texttt{SB2},  \texttt{EB}   \\
462130376453494272  & 136 & 0.0 &  \texttt{EB}, \texttt{SB9}   \\
5329693847558732288  & 135 & 0.0 &  \texttt{SB2},  \texttt{EB}   \\
5966509022176616448  & 123 & 0.0 &  \texttt{SB2}   \\
2006888825592397440  & 122 & 0.0 & \texttt{EB}   \\
4094832952561701888  & 108 & 0.0 &  \texttt{SB2} \\ 
5527324946602623360  & 108 & 0.0 &  \texttt{SB2},  \texttt{EB}   \\
5858981908977997312  & 105 & 0.0 &  \texttt{EB}   \\
5619903482072419712  & 105 & 0.8 &     \\
5999680605109166848  & 96 & 0.0 &  \texttt{SB2},  \texttt{EB}   \\
4043375469172364288  & 93 & 0.0 &  \texttt{EB}   \\
5337747731560719616  & 91 & 0.0 & literature (HD\,96670)    \\
4066063754109686016  & 91 & 0.0 & \texttt{EB}, literature (HD\,165246)\\
4156190653517727360  & 90 & 0.0 & \texttt{SB9}   \\
5868478390840987776  & 89 & 0.0 &  \texttt{SB2}   \\
2059011376158159488  & 87 & 0.0 &  \texttt{EB}, \texttt{SB9}  \\
5966522667303699072  & 87 & 0.0 &  \texttt{SB2},  \texttt{EB}   \\
3113547759973829120  & 86 & 0.0 & \texttt{LPV}    \\
5350350303799853056  & 85 & 0.0 &  \texttt{SB2},  \texttt{EB}   \\
5979913172849957504  & 84 & 0.0 &  \texttt{SB2}   \\
2060647728648558464  & 84 & 0.0 & \texttt{SB9}    \\
5970357660807157632  & 83 & 0.0 & literature (HD\,153426)  \\
528570015826682496  & 83 & 0.0 &  \texttt{SB2}   \\
5963159364309081472  & 83 & 0.0 &  \texttt{EB}   \\
3108642147107724544  & 80 & 0.0 &  \texttt{EB}   \\
5966509541883497984  & 80 & 0.0 & \texttt{SB9}   \\
5968081461246595456  & 79 & 0.0 &  \texttt{EB}   \\
5969671118856579072  & 76 & 0.0 &  \texttt{SB2}   \\
4069547797210386816  & 75 & 0.0 & literature (HD\,163892)  \\
5966450786728068736  & 74 & 0.8 &    \\
5351446924882928128  & 74 & 0.8 &    \\
5966563486675142912  & 73 & 0.0 & literature (HD\,152408)   \\
6057653936254670464  & 72 & 0.0 &  \texttt{SB2}    \\
6055854065741096960  & 69 & 0.0 &  \texttt{EB}   \\
5350347108343919104  & 67 & 0.8 &    \\
5864726788439265024  & 66 & 0.0 & \texttt{Be star} \\
5968259964433588992  & 64 & 0.0 &  \texttt{SB2}   \\
5863473757493410688  & 64 & 0.8 &    \\
5862482685215601664  & 64 & 0.0 &  \texttt{SB2}   \\
4096387528961643136  & 64 & 0.0 & \texttt{Be star}   \\
431784233486202880  & 62 & 0.8 &    \\
5975565497602725376  & 60 & 0.8 &    \\
6056407639825640448  & 60 & 0.0 &  \texttt{SB2},  \texttt{EB}   \\
5859147488560637696  & 60 & 0.0 &  \texttt{EB}   \\
4044071494394682752  & 59 & 0.0 & \texttt{Be star}   \\
5977135497163587712  & 58 & 0.8 &    \\
5977345710074220800  & 58 & 0.0 & literature (HD\,154643)  \\
5969686958696979712  & 57 & 0.0 & literature (HD\,152405)  \\
5350391535488119552  & 56 & 0.0 &  \texttt{SB2}, \texttt{SB9}   \\
5884713092342630144  & 55 & 0.0 & \texttt{LPV}   \\
5522306019626566528  & 55 & 0.0 & literature (HD\,74194)  \\
182722859557158272  & 54 & 0.8 &   \\ 
2060804993182977664  & 54 & 0.8 &  \\ 
4054675493689626240  & 53 & 0.0 &  \texttt{SB2},  \texttt{EB}   \\
\end{supertabular}
\end{center}

The resulting classifications for all 66 stars are given in Table~\ref{tab:follow_up}.
We note that the detection of a luminous secondary excludes a BH as the close, RV-inducing companion, but does not necessarily exclude a BH elsewhere in a higher-order system. Therefore, the quantity constrained in this work is the fraction of OB stars whose close, RV-inducing companion is a BH, rather than the fraction of OB stars containing a BH in any configuration.

\section{Forward-model of RV amplitudes}
\label{app:bh_model}

\begin{figure}[h]
\centering
\includegraphics[width=0.9\columnwidth]{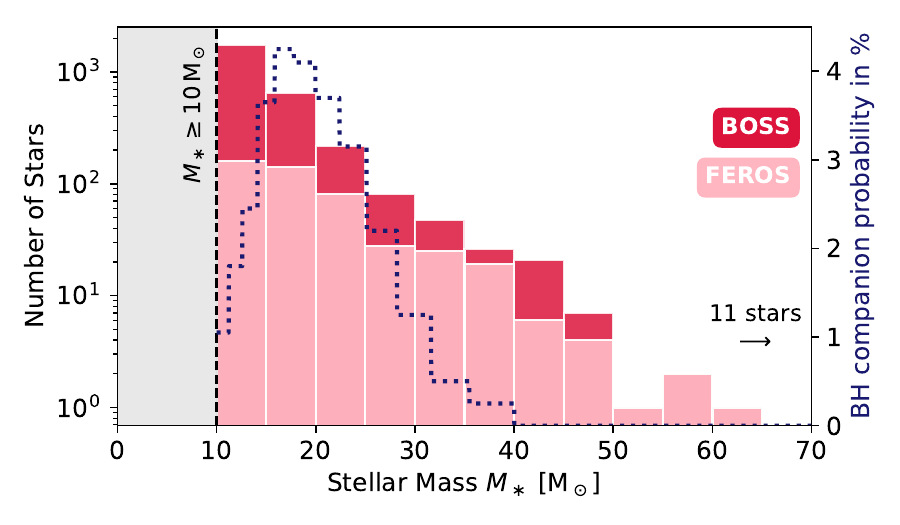}
\caption{Distribution of inferred stellar masses in the OB parent sample. The sample is limited to massive stars with inferred $M_\mathrm{OB} {\geq} 10\,\mathrm{M}_\odot$. The dotted blue line shows the mass-dependent probability for an OB star to have a BH companion as predicted by the \citet{Langer+2020} population synthesis models, peaking for stellar masses between $M_\mathrm{OB} {\simeq} 15$--$20\,\mathrm{M}_\odot$.
 }
\label{fig:stellar_masses_histogram}
\end{figure}

To assess the sensitivity of our RV survey to OB+BH binaries and estimate the detection probabilities $P_{\mathrm{det},i}^\mathrm{BH}$ used in the hierarchical inference, we forward-modelled the expected RV amplitudes for the stars in our parent sample. We adopted the BH companion properties predicted by
\citet{Langer+2020} and simulated the RV signal at the actual observing epochs of each star. This allowed us to account explicitly for the different temporal baselines and cadence of the BOSS and FEROS observations.

\subsection{Adopted BH population}

We adopted the BH mass and orbital-period distributions from \citet{Langer+2020}. Since the full joint probability distributions of the binary parameters are not available, we reconstructed the relevant marginal distributions from the distributions shown in their figures and sampled the parameters independently. This approximation neglects possible correlations between parameters, such as orbital period and stellar or BH mass. The resulting detection probabilities are therefore conditional on the adopted marginal distributions rather than a full reproduction of the population-synthesis model.

The BH mass distribution was reconstructed from the mass distribution shown in Fig.~5 of \citet{Langer+2020}. We divided the range $6.6$--$20\,M_\odot$ into eight log-uniformly spaced bins and assigned the relative probabilities reported by the model to these bins. The resulting distribution was then normalised and sampled as a piecewise-constant probability density within the bin boundaries. 

For the orbital period, we reconstructed the distribution shown in the upper panel of Fig.~6 of \citet{Langer+2020} as a three-component mixture in $\log_{10}(P_\mathrm{orb}/{\rm d})$. The dominant component, corresponding to Case~B systems, was described by an asymmetric normal distribution centred at $\log_{10}(P_\mathrm{orb}/{\rm d}) {=} 2.25$, with standard deviations $\sigma_{\rm lo} {=} 0.70$ and $\sigma_{\rm hi} {=} 0.35$ on the two sides of the peak, and has a relative weight of 0.89. We additionally included the short-period Case~A component centred at $\log_{10}(P_\mathrm{orb}/{\rm d}) {=} 0.75$, with $\sigma_{\rm lo} {=} 0.10$ and $\sigma_{\rm hi} {=} 0.16$, and a relative weight of 0.105, as well as the small non-interacting tail centred at $\log_{10}(P_\mathrm{orb}/{\rm d}) {=} 3.55$, with $\sigma_{\rm lo} {=} 0.05$ and $\sigma_{\rm hi} {=} 0.01$, and a relative weight of 0.015. The three weights were normalised to sum to unity. The resulting period distribution is therefore dominated by systems
with periods of order $10^2$~d, with additional contributions from short-period and long-period systems.

We assumed circular orbits, as in the Langer et al.~models, and adopted the corresponding assumption of zero natal kicks. Orbital inclinations were drawn isotropically, that is, uniformly in $\cos i$, while the argument of periastron was drawn uniformly between $0$ and $2\pi$. The orbital phase was randomised by drawing the time of periastron uniformly between zero and one orbital period. For each star, we then calculated the RV semi-amplitude from its inferred stellar mass, the simulated BH mass, orbital period, and inclination.

\subsection{Detection probability}

To determine the probability of detecting such systems in our RV survey, we assigned a simulated BH companion to every star in the parent sample and generated a large number of independent binary realisations. For each realisation, we calculated the expected RV curve at the actual observation times of the corresponding star and the maximum observable RV difference $\Delta {\rm RV}$. As for the observations, we defined RV-detectable systems following Eq.~\eqref{eq:rv_criterion}. For each star, the detection probability $P_{\mathrm{det},i}^\mathrm{BH}$ was taken as the fraction of simulated BH realisations satisfying this criterion.
In other words, we Monte Carlo sample assumed distributions for the BH mass ($M_\mathrm{BH}$) and the orbital period ($P_\mathrm{orb})$ to estimate
\begin{multline*}
    P_{\mathrm{det},i}^\mathrm{BH} \equiv \int\limits_{50\mathrm{km/s}}^\infty d(\Delta \mathrm{RV}) \\
    \int dM_\mathrm{BH} dP_\mathrm{orb} \, p(M_\mathrm{BH},P_\mathrm{orb}) p(\Delta RV|M_{\mathrm{OB},i}, M_\mathrm{BH}, P_\mathrm{orb})
\end{multline*}
where $p(M_\mathrm{BH}, P_\mathrm{orb})$ represents the mass and period distribution from~\citet{Langer+2020}.
Across the parent sample, $P_{\mathrm{det}, i}^\mathrm{BH}$ has a median of approximately 39\%. The star-to-star variation is driven primarily by the number of epochs and temporal baseline of the RV observations.

\subsection{Mass-dependent BH fraction}

The Langer et al.~models also predict a mass-dependent probability for an OB star to have a BH companion. We reconstructed this probability from their Fig.~11 using bins of width $0.05$ in $\log_{10}(M_\mathrm{OB}/M_\odot)$, see Fig.~\ref{fig:stellar_masses_histogram}. The resulting BH probability rises from approximately $1\%$ at $M_\mathrm{OB} {\simeq} 10\,\mathrm{M}_\odot$ to a maximum of about $4\%$ around $M_\mathrm{OB} {\simeq} 15$--$20\,M_\odot$, and decreases towards higher masses. We set the probability to zero outside the $10$--$40\,\mathrm{M}_\odot$ range covered by the model.
We used these mass-dependent probabilities to estimate the expected BH fraction for our observed OB-star mass distribution. Specifically, we evaluated the Langer et al.~BH probability for each star in the parent sample and took the mean over the sample. This gives a BH companion fraction of approximately $2.5\%$ for our observed stellar-mass distribution. This value assumes the same binary-fraction normalisation as \citet{Langer+2020}, namely that the model is normalised to an initial binary fraction of $100\%$.

\subsection{Model assumptions and limitations}

The forward model is necessarily conditional on the population model adopted from \citet{Langer+2020}. We note that reconstructing the population from marginal distributions neglects correlations between binary parameters that may be present in the underlying population-synthesis calculation. The Langer et al.~models also assume an initial binary fraction of 100\%, so their predicted OB+BH occurrence fractions scale approximately linearly with the adopted binary fraction. In the adopted supernova model, BHs are formed by direct collapse from stars with helium core mass above $6.6\,\mathrm{M}_\odot$ at the time of core carbon exhaustion, which is a comparably optimistic BH formation criterion \citep[see e.g.][]{Maltsev+2025}. 

A further caveat is that the Langer et al.~models were calculated at subsolar metallicity, representative of the Large Magellanic Cloud. We expect their qualitative results to remain applicable at approximately solar metallicity, with the largest expected difference being in the BH mass distribution \citep{Janssens+2024}. The stronger stellar winds at higher metallicity can lead to greater mass loss and, consequently, lower remnant masses. On average, lower BH masses would result in smaller RV amplitudes and hence somewhat lower detection probabilities. We tested the effect of lower BH masses in Appendix~\ref{app:robustness}.

The Langer et al.~models also assume zero natal kicks. 
Natal kicks may affect both the number of surviving OB+BH systems and their orbital properties. In particular, kicks may disrupt a fraction of the binaries and can introduce orbital eccentricity, thereby affecting the predicted RV detection probabilities. Since our aim is to derive an upper limit, we retain the Langer et al.~population model as our reference model, including its zero-kick assumption, and interpret the resulting constraint as model-dependent. We tested the effect of eccentric orbits in Appendix~\ref{app:robustness}.

Finally, the adopted period distribution does not constrain populations of substantially wider OB+BH binaries. \citet{Langer+2020} note that non-interacting systems may form with orbital periods exceeding approximately $3\,\mathrm{yr}$, but their models are restricted to shorter periods by the upper initial-period limit of their model grid. Such wide systems are also more susceptible to disruption by even small BH formation kicks and would be particularly difficult to detect in our RV data, either because the induced RV variations are small or because the orbital period is long compared to our temporal baseline. We therefore focus on close to intermediate-period OB+BH systems. Our inferred limit should consequently be interpreted as a constraint on the population with orbital properties represented by the adopted Langer et al.~model, rather than on all OB stars that may host a BH at any orbital separation.

\section{Derivation of the follow-up likelihood}
\label{app:follow_up-likelihood}

For each RV-variable star $j$ with follow-up information, we define
$\phi_j \equiv P({\rm BH}\mid D_j, p_{\rm ref})$,
where $D_j$ denotes the follow-up data and $p_{\rm ref}$ is a reference
prior that assigns equal probability to the BH and non-BH hypotheses.
Thus, $\phi_j$ describes the evidence provided by the follow-up data for
star $j$ when considered in isolation, before incorporating any information
about the population occurrence fraction.

In the hierarchical model, however, the prior probability that an
RV-variable star $j$ hosts a BH is determined by the population model,
\begin{equation}
\pi_{{\rm BH},j}
=
P({\rm BH}\mid Y_j {=} 1,f_{\rm BH},\kappa_s)
=
\frac{f_{\rm BH}P_{{\rm det},j}^{\rm BH}}
{f_{\rm BH}P_{{\rm det},j}^{\rm BH}+\kappa_{s(j)}}.
\end{equation}

\noindent The follow-up likelihood must therefore account for the probability of the
observed data under both possible companion classes,
\begin{equation}
P(D_j\mid \pi_{{\rm BH},j})
=
P(D_j\mid{\rm BH})\pi_{{\rm BH},j}
+
P(D_j\mid\neg{\rm BH})(1-\pi_{{\rm BH},j}).
\label{eq:followup_marginal}
\end{equation}

\noindent The quantities required in Eq.~\eqref{eq:followup_marginal} are likelihoods, whereas $\phi_j$ is a posterior probability obtained under the reference prior. We therefore recover the likelihood ratio implied by $\phi_j$.
Applying Bayes' theorem with equal prior probabilities gives
\begin{equation}
\phi_j = \frac{P(D_j\mid{\rm BH})} {P(D_j\mid{\rm BH})+P(D_j\mid\neg{\rm BH})} \Longrightarrow \frac{P(D_j\mid{\rm BH})} {P(D_j\mid\neg{\rm BH})} = \frac{\phi_j}{1-\phi_j}.
\label{eq:likelihood_ratio}
\end{equation}

\noindent Only this likelihood ratio is relevant for the hierarchical inference: multiplying both likelihoods by the same factor changes the total likelihood only by a constant that is independent of the model parameters.
We can therefore adopt the convenient normalisation
$P(D_j\mid{\rm BH}) \propto \phi_j$ and 
$P(D_j\mid\neg{\rm BH}) \propto 1-\phi_j$.

\noindent Substituting these expressions into Eq.~\eqref{eq:followup_marginal}, the contribution of star $j$ to the follow-up likelihood is, up to a multiplicative constant,
\begin{equation}
P(D_j | Y_j {=} 1,f_\mathrm{BH},\kappa_s) \propto \pi_{{\rm BH},j}\phi_j + \left(1-\pi_{{\rm BH},j}\right) \left(1-\phi_j\right).
\end{equation}

\noindent This expression has the expected limiting behaviour. If $\phi_j {=} 0$, the follow-up data exclude the BH hypothesis and the likelihood contribution becomes  $1 {-} \pi_{{\rm BH},j}.$
Conversely, $\phi_j {=} 1$ corresponds to a definite BH identification and gives $P(D_j|Y_j {=} 1,f_\mathrm{BH},\kappa_s) {=} \pi_{{\rm BH},j}.$
Finally, for $\phi_j {=} 0.5$, $P(D_j|Y_j {=} 1,f_\mathrm{BH},\kappa_s) {=} 0.5,$ which is independent of the model parameters and therefore provides no information to the inference. 

\section{Robustness checks}
\label{app:robustness}

We performed a series of robustness checks to assess the sensitivity of our inferred upper limit on $f_\mathrm{BH}$ to the assumptions entering both the inference model and the adopted OB+BH population. The results are summarised in Table~\ref{tab:robustness_checks}.

\setlength{\tabcolsep}{10pt}
\begin{table}[h]
\centering
\caption{Results of the robustness tests. The table lists the modification relative to the fiducial model and the corresponding inferred 95\% upper limit on $f_\mathrm{BH}$.}
\label{tab:robustness_checks}
\begin{tabular}{lr}
\toprule
Model  
& $f_{\rm BH}^{95}$\\
\midrule
{fiducial model} & \textbf{$\leq 3.19\%$}  \\
\midrule
\addlinespace
with separate $\kappa_{s, \mathrm{FEROS}}, \, \kappa_{s, \mathrm{BOSS}}$ & $\leq 3.08\%$\\
\addlinespace
marginally uniform $f_\mathrm{BH}$ prior  & $\leq 3.21\%$  \\
log-uniform $f_\mathrm{BH}$ prior & $\leq 1.40\%$ \\
\addlinespace
RV cut-off $\Delta\mathrm{RV}>35$ km\,s$^{-1}$ & $\leq 3.03\%$  \\
RV cut-off $\Delta\mathrm{RV}>65$ km\,s$^{-1}$& $\leq 2.29\%$ \\
\addlinespace
with empirical $P_\mathrm{det, bin}$ instead of $\kappa_s$& $\leq 2.93\%$  \\
\addlinespace
with eccentric OB+BH orbits & $\leq 3.65\%$ \\ 
with lowered BH masses & $\leq 3.22\%$ \\ 
\bottomrule
\end{tabular}
\tablefoot{$\epsilon$ refers to the fraction of RV-variable non-BH systems that escape SB2 detection, see Appendix~\ref{app:follow_up}. The value adopted in the fiducial model is $\epsilon=0.235$.}
\end{table}
\setlength{\tabcolsep}{6pt}

\subsection{Luminous-binary contamination} 
In the fiducial inference, the contribution of stellar binaries to the RV-variable population is described by the effective nuisance parameters $\kappa_s$. 
As a test for possible differences between the BOSS and FEROS samples, we relaxed the assumption that the luminous-binary contribution is the same in the two data sets. We therefore repeated the inference with independent $\kappa_{s,\mathrm{BOSS}}$ and $\kappa_{s,\mathrm{FEROS}}$ for each cadence bin. The resulting posterior on $f_\mathrm{BH}$ and its 95\% upper limit are consistent with the fiducial result, with the upper limit remaining at $3.08\%$. This indicates that allowing for different luminous-binary contamination in the BOSS and FEROS subsamples does not substantially change our constraint.

\subsection{Prior on $f_\mathrm{BH}$} 
The fiducial model adopts a prior that is uniform over the full physically allowed parameter volume in $(f_\mathrm{BH},\kappa_1,\kappa_2,\kappa_3)$. We tested two alternative prior choices. First, we adopted a prior that is uniform in the marginal distribution of $f_\mathrm{BH}$, while retaining uniform conditional priors for the nuisance parameters, $0 {\leq} \kappa_s {\leq} (1 {-} f_\mathrm{BH})$. Second, we adopted a log-uniform prior on $f_\mathrm{BH}$ in the range $10^{-4}$ to $10^0$. The marginally uniform prior yields the same 95\% upper limit as the fiducial model of $3.21\%$. The log-uniform prior results in a tighter upper limit of $1.40\%$. 

\subsection{RV variability threshold} 
The fiducial RV threshold of $\Delta\mathrm{RV} {>} 50\,\mathrm{km\,s^{-1}}$ was chosen to separate the majority of apparently single stars and low-amplitude variables from the higher-amplitude RV-variable population, while retaining a substantial fraction of the expected OB+BH systems and high follow-up classification completeness. The precise value is nevertheless somewhat arbitrary. We therefore repeated the analysis using thresholds of $35$ and $65\,\mathrm{km\,s^{-1}}$. The resulting 95\% upper limits are $3.03\%$ and $2.29\%$, respectively, compared to $3.19\%$ for the fiducial threshold. Our results are therefore only weakly sensitive to the choice of RV threshold. 

\subsection{Alternative OB+BH population models} 
Our results are model-dependent and should be interpreted in the context of the OB+BH parameter distributions adopted from \citet{Langer+2020}. A comprehensive comparison of different population-synthesis models is beyond the scope of this work. We nevertheless tested two modifications motivated by the limitations discussed in Appendix~\ref{app:bh_model}: lower BH masses and non-zero orbital eccentricities. First, we replaced the fiducial BH mass distribution with a modified more Milky-Way-appropriate distribution, roughly reconstructing the distribution from \citet{Janssens+2024}, their Fig.~B.1, which accounts for the expected lower remnant masses at higher metallicity. Second, we replaced the circular orbits of the fiducial model with a uniform eccentricity distribution to assess the effect of non-zero eccentricities. For both modifications, we repeated the forward modelling of the RV detection probabilities and the subsequent Bayesian inference following the procedure described in Appendix~\ref{app:bh_model}. The resulting 95\% upper limits are $3.22\%$ for the lower BH masses and $3.65\%$ for the eccentric-orbit model. Both remain close to the fiducial value, indicating that these assumptions do not strongly affect our inferred constraint.

\subsection{Explicit modelling of the luminous-binary population} 
\label{app:robustness_fbin-model}
Finally, we tested the approximation of describing the luminous-binary contamination using the three nuisance parameters $\kappa_s$. This approach has the advantage of avoiding a detailed model of the binary orbital-parameter distributions and their detection probabilities, while remaining flexible enough to describe any population contributing to the RV-variable sample. However, it does not explicitly account for the star-to-star variation in the probability of detecting a luminous binary, which depends on the orbital parameters, number of epochs, and temporal sampling. The binary fraction and orbital-parameter distributions of the SDSS-V OB-star sample are not yet sufficiently constrained for a detailed population model; their inference is the subject of separate work (Villase\~nor et al., in prep.). As an approximate test of the effect of explicitly modelling the luminous-binary population, we replaced the $\kappa_s$ parameters with a binary fraction $f_\mathrm{bin}$ and computed a per-star detection probability $P_{\mathrm{det}}^\mathrm{bin}$, analogous to the treatment of the OB+BH population. We adopted empirical distributions of binary parameters for massive O-type stars based on results by \citet{Sana+2012}. Specifically, we sampled orbital periods from a power-law distribution in $\log P_\mathrm{orb}$ with power-law index $\alpha {=} {-}0.5$, eccentricities from a power law with index $\eta {=} {-}0.4$, and mass ratios from a uniform distribution between 0.1 and 1. We repeated the forward modelling and Bayesian inference using this alternative model architecture and obtained a 95\% upper limit of $2.93\%$. Although a more detailed treatment of the luminous-binary population could refine this estimate in future work, the approximate agreement with the fiducial result provides a useful check that the simplified $\kappa_s$ parametrisation does not dominate our inferred constraint.

Across all tests considered here, the 95\% upper limit ranges from $1.40\%$ to $3.65\%$. The largest changes arise from replacing the circular-orbit assumption with a uniform eccentricity distribution for
the OB+BH systems. Given this scatter, we estimate the systematic uncertainty on the 95\% upper limit to be of order $1$ percentage point and therefore report our main result as $f_{\rm BH}\lesssim(3\pm1)\%$.

\section{Scaling of the OB+BH constraint with follow-up completeness}
\label{app:experiment_design}

Section~\ref{sec:discussion_strategy} argues that the constraint on $f_{\rm BH}$ is primarily set by the follow-up rather than by the size of the parent sample. Here we illustrate that scaling in a simplified limit.

We consider $m_{\rm fu}$ RV-variable stars with follow-up information: $k$ identified as OB+BH systems ($\phi_j {=} 1$), $d$ classified as non-BH ($\phi_j {=} 0$), and $u$ inconclusive, for which no companion is detected and $\phi_j {=} 1/(1+\epsilon)$ as in Appendix~\ref{app:follow_up}. If the followed-up stars form a relatively homogeneous subsample, their class and detection probabilities can be approximated by common values $\pi_{\rm BH}$ and $P_{\rm det, fu}^{\rm BH}$, and the
follow-up likelihood (Eq.~\ref{eq:likelihood_fu}) becomes
\begin{equation}
p(\{D_j\}|\{Y_j {=} 1\},f_\mathrm{BH},\kappa_s)
\propto
\pi_{\rm BH}^{\,k}\,
(1-\pi_{\rm BH})^{d}\,
\bigl[\epsilon+\pi_{\rm BH}(1-\epsilon)\bigr]^{u}.
\label{eq:binomial_followup}
\end{equation}
The hierarchical model, in Eq.~\eqref{eq:qbh}, relates $\pi_{\rm BH}$ to the occurrence fraction.
A constraint on $\pi_{\rm BH}$ therefore translates directly into one on $f_{\rm BH}$.

\paragraph{Detection regime.}
If OB+BH systems are identified ($k {>} 0$) and all classifications are decisive ($u {=} 0$), Eq.~\eqref{eq:binomial_followup} is binomial. Maximising the likelihood gives $\hat \pi_{\rm BH} {=} k/m_{\rm fu}$ and a fractional precision
\begin{equation}
\frac{\sigma_{f_{\rm BH}}}{\hat f_{\rm BH}}
\approx
\sqrt{\frac{1-\hat \pi_{\rm BH}}{\hat \pi_{\rm BH}\,m_{\rm fu}}}
\approx
\frac{1}{\sqrt{k}},
\qquad \hat \pi_{\rm BH}\ll1 .
\end{equation}
Thus, once OB+BH systems are detected, the fractional precision of $f_{\rm BH}$ is set approximately by the number of detections and scales as $k^{-1/2}$.
\paragraph{Zero-detection regime.}
The present dataset instead has $k {=} 0$. For $\pi_{\rm BH}\ll1$, expanding the
logarithm of Eq.~\eqref{eq:binomial_followup} to first order gives
\begin{equation}
P(\{D_j\}|\{Y_j {=} 1\},f_\mathrm{BH},\kappa_s)
\propto
\exp\left(-m_{\rm fu,eff}\,\pi_{\rm BH}\right),
\end{equation}
where
\begin{equation}
m_{\rm fu,eff}
=
d-u\frac{1-\epsilon}{\epsilon}
=
\sum_j\frac{1-2\phi_j}{1-\phi_j}.
\label{eq:meff}
\end{equation}
The second expression follows from the same expansion applied to a general $\phi_j$.
To this effective number of follow-up stars $m_{\rm fu,eff}$, each follow-up target contributes according to the information in its classification: $+1$ for a decisive non-BH classification, zero for $\phi_j {=} 0.5$, and ${-} (1 {-} \epsilon)/\epsilon$ for an inconclusive non-detection, because it is more likely under the BH hypothesis than under the non-BH hypothesis.

For an approximately exponential likelihood, taking a likelihood ratio of $0.05$ as the characteristic 95\% constraint gives $\pi_{\rm BH}^{95} {\sim} {-} \ln(0.05)/m_{\rm fu,eff} {\sim} 3/m_{\rm fu,eff}$ and,
with Eq.~\eqref{eq:qbh},
\begin{equation}
f_{\rm BH}
\simeq
\pi_{{\rm BH}}\,
\frac{n_\mathrm{RV} / N}{P_{\rm det, fu}^{\rm BH}}
\Rightarrow
f_{\rm BH}^{95}
\sim
\frac{3}{m_{\rm fu,eff}} \frac{n_\mathrm{RV} / N}{P_{\rm det, fu}^{\rm BH}},
\label{eq:fbh95}
\end{equation}
where we approximated $P(Y_j{=}1|f_\mathrm{BH},\kappa_s,M_{\mathrm{OB},j}) \sim n_\mathrm{RV} / N$ as a representative value for the sample. So in the zero-detection regime, the upper limit improves approximately as $m_{\rm fu,eff}^{-1}$. 

For incomplete classification and a fixed number of systems that were followed up, inconclusive non-detections reduce the effective number of informative follow-ups and therefore weaken the constraint.
This means that sensitivity depends not only on the number of followed-up targets but also on classification completeness. Adding targets without increasing the number of decisive classifications does not necessarily improve the constraint. For example, archival cross-matches of the full RV-variable sample would add some exclusions but many more inconclusive cases. 
This would also likely increase $\epsilon$ substantially with the result that $m_\mathrm{fu,eff}$ may not change significantly, consistent with the intuition that adding more data should not weaken constraints.
There is clearly a balance between adding a few additional systems with conclusive follow-up vs. adding many systems with inconclusive follow-up, and the optimal observing strategy may depend on the details of how much each type of follow-up costs.
This balance motivated concentrating the follow-up on the subset with high-resolution spectra.
For the present sample, $d {=} 56$, $u {=} 10$, and
$\epsilon {=} 0.235$ give $m_{\rm fu,eff} {\approx} 23$, so that with
$n_\mathrm{RV} / N {\approx} 0.13$ and $P_{\rm det, fu}^{\rm BH} {\approx} 0.47$,
Eq.~\eqref{eq:fbh95} yields $f_{\rm BH}^{95} {\sim} 3.5\%$, 
close to the full hierarchical constraint.

\end{appendix}
\end{document}